\documentclass{JHEP3}
\usepackage[dvips]{graphicx}
\usepackage{latexsym,amsmath,amssymb}

\newcommand{\cpl}[3]{Chin.\ Phys.\ Lett.\ \textbf{#1} (#2) #3}

\title{Thermodynamics of a black hole based on a
generalized uncertainty principle}

\author{Wontae Kim \\
  Department of Physics and Center for Quantum Spacetime, Sogang University,
  C.P.O. Box 1142, Seoul 100-611, Korea \\
  E-mail: \email{wtkim@sogang.ac.kr}}
\author{Edwin J. Son \\
  Department of Physics and Basic Science Research Institute, Sogang
  University, Seoul 121-742, Korea \\ 
  E-mail: \email{eddy@sogang.ac.kr}}
\author{Myungseok Yoon \\ 
  Center for Quantum Spacetime, Sogang University, Seoul 121-742,
  Korea \\
  E-mail: \email{younms@sogang.ac.kr}}

\date{\today}

\abstract{%
We study thermodynamic quantities and the stability of a black hole in
a cavity using the Euclidean action formalism by Gibbons and Hawking
based on the generalized uncertainty relation which is extended in a
symmetric way with respect to the space and momentum without loss of
generality. Two parameters in the uncertainty relation affect the
thermodynamical quantities such as energy, entropy, and the heat
capacity. In particular, it can be shown that the small black hole is
unstable and it may decay either into a minimal black hole or a large
black hole. We discuss a constraint for a large
black hole comparable to the size of the cavity in connection with the
critical mass.}


\begin{document}

\section{Introduction}
\label{sec:intro}

Hawking's quantum field theoretical analysis~\cite{hawking} has shown that
a Schwarzschild black hole has a thermal radiation with a temperature
$T_{\rm H} = (8\pi M)^{-1}$, where $M$ is the mass of the black
hole. Subsequently, this issue has been investigated in the
thermodynamical regime through the path-integral approach to the
quantization of gravity~\cite{gh,hi}. It has been also shown that the
entropy of a black hole is always equal to one quarter of the area of the
event horizon in fundamental units and a stationary system without
event horizon has no entropy. Moreover, the thermodynamics in an
asymptotically anti-de Sitter black hole has been studied in
Ref.~\cite{bcm}. Using the Euclidean action approach by Gibbons and
Hawking~\cite{gh}, the thermodynamic local quantities such as
temperature, energy, entropy, and surface pressure, have been
evaluated in a cavity with a finite size~\cite{allen,york,wy,brown}. 
Unlike other quantities related to the
size of cavity, the entropy does not have local property of gravity
since it agrees with Bekenstein-Hawking entropy depending only 
on the event horizon. It means that the entropy is independent of the
asymptotic behavior of fields.

Now, the conventional Heisenberg uncertainty
principle(HUP) has been promoted to the generalized uncertainty
principle(GUP)~\cite{pad,maggiore,garay,kmm,kln} based on some aspects of
quantum gravity and the string theory, which is given by 
\begin{equation}
  \Delta x \Delta p \ge \hbar \left( 1 +
      \ell^2 \frac{(\Delta p)^2}{\hbar^2} \right), \label{ogup}
\end{equation}
where it leads to the minimal length of $\Delta x_{\rm min} =
2\ell$. The cutoff $\ell$ may be chosen as a string scale in the
context of the perturbative string theory or Plank scale based on the
quantum gravity. In the brick-wall method~\cite{thooft}, the GUP has
been used to calculate the entropy of black
holes without a cutoff parameter~\cite{li,liu,kkp,yhk} where the
minimal length plays the role of ultraviolet cutoff and it is regarded
as a natural cutoff. Also, the corrections to entropy by the GUP has been studied in other methods~\cite{mv}. Recently, the thermodynamics and its stability
for the Schwarzschild black hole have been studied by applying the
GUP~\cite{acs,ch,mkp}. They obtained a remnant after evaporation of a
black hole and it may be stable, however, the relevant thermodynamic
quantities should be treated as local quantities because the GUP
effects significantly appear near horizon.

On the other hand, one can generalize the GUP by considering $(\Delta
x)^2 $ along with $(\Delta p)^2$ in the uncertainty relation~(\ref{ogup})
for the same footing~\cite{hk}. Then, the symmetric generalized uncertainty
principle(SGUP) can be written by
\begin{equation}
  \Delta x \Delta p \ge \hbar \left( 1 + \frac{(\Delta x)^2}{L^2} +
      \ell^2 \frac{(\Delta p)^2}{\hbar^2} \right), \label{gup}
\end{equation}
which leads to the minimal length of $\Delta x_{\rm min} =
2\ell/\sqrt{1-4\ell^2/L^2}$ and the minimal momentum of $\Delta p_{\rm
  min} = 2\hbar/(L\sqrt{1-4\ell^2/L^2})$, where $L$ is
another uncertainty constant.

In this work, based on the SGUP, we would like to study the
thermodynamic behaviors of physical quantities of a black hole in
the Euclidean action formalism which gives the local Tolman 
temperature naturally, and investigate the stability of the black hole in
terms of the heat capacity. When the temperature is over the critical
temperature~\cite{gpy,hp,sh}, the small black hole created by the
phase transition is unstable and decays to hot flat space or grows to
the cavity size, which can be expanded to the infinity. However, in
the GUP improved thermodynamics, the small black hole cannot decay to
hot flat space by the thermal radiation since there is the minimal
size of a black hole. The difference from the previous
works~\cite{acs,ch,mkp} mainly comes from the local Tolman temperature
whereas the global temperature has been used in the thermodynamic
analysis so far. So, in Sec.~\ref{sec:gup}, we shall obtain the local
Tolman temperature in this SGUP and then calculate the thermodynamic
local quantities compatible with the local temperature related to the
size of a cavity. The local entropy, which is consistent with the
thermodynamic first law, will be also derived. In
Sec.~\ref{sec:largeL}, we will take the limit of $L\to\infty$ called
the GUP case and investigate the thermodynamics and the stability of
the black hole which have not been discussed in earlier works. In
Sec.~\ref{sec:comp}, when $L$ is finite, thermodynamic analysis will
be done. Finally, we draw some discussions in Sec.~\ref{sec:dis}.

\section{Thermodynamic quantities in SGUP}
\label{sec:gup}
The thermodynamic quantities will be defined in a cavity, which
means that we have to consider the local temperature based on the SGUP. 
It seems to be plausible to consider the local temperature rather than
the Hawking temperature in the cavity. 
For this purpose, we solve Eq.~(\ref{gup}) for the momentum uncertainty
in terms of the position uncertainty~\cite{acs} 
\begin{equation}
  \frac{\Delta p}{\hbar} = \frac{\Delta x}{2\ell^2} \left( 1 \pm
    \sqrt{1 - \frac{4\ell^2}{L^2} - \frac{4\ell^2}{(\Delta x)^2}}
  \right), \label{mom}
\end{equation}
and putting $\Delta x = 2M$, we identify the emitted photon energy with
the black hole temperature up to a calibration factor so that
\begin{equation}
  T_{SGUP} = \frac{M}{4\pi\ell^2} \left( 1 - \sqrt{1 -
  \frac{4\ell^2}{L^2} - \frac{\ell^2}{M^2}} \right)
  , \label{T*}
\end{equation}
where we set $\hbar=G=1$ for simplicity.
We assume on the basis of thermodynamic consistency that the emitted photons have a thermal black body spectrum. The remainder of our work depends upon this assumption.
If we consider a large black
hole($\ell/M \ll 1$) with $M \ll L$, then the modified temperature, in
the leading order, goes to the well-known Hawking temperature when we
choose the negative sign. So, the limiting case of $\ell \rightarrow
0$ and $L \to \infty$ is well-defined. If we set $L\lesssim M$, then it
gives a correction to the Hawking temperature, $T_{SGUP}
\approx 1/(8 \pi M) + M/ (2\pi L^2)$.
   
Now, the partition function can be written as $Z = \exp(-I) =
\exp(-\beta F)$, where $I$, $\beta$, and $F$ are the first-order
Euclidean Einstein action, the inverse of temperature $T$, and the
free energy of system in the cavity of the finite radius. The
Euclidean action with a subtraction term is defined by
\begin{equation}
  I = I_1 - I_{\rm 0}, \label{I}
\end{equation}
where
\begin{equation}
  I_1 = -\frac{1}{16\pi} \int d^4 x \sqrt{g}\, {\cal R} +
  \frac{1}{8\pi} \oint d^3 x \sqrt{\gamma} K. \label{I1}
\end{equation}
Here, $K$ is the trace of the extrinsic curvature tensor $K_{ij}$ of
the boundary $S^1 \times S^2$ of $r= R = {\rm const.}$ and
$\gamma_{ij}$ is its induced three-metric. The line element of
a Schwarzschild black hole is
\begin{equation}
  ds_{\rm E}^2 = f d\tau^2 + f^{-1} dr^2 + r^2 d\theta^2 + r^2
  \sin^2\theta^2 d\phi^2, \label{metric}
\end{equation}
where $f(r) = 1-2M/r$ and $\tau$ is the Euclidean time. The period of
the Euclidean time is $\beta_{SGUP} = T_{SGUP}^{-1}$, and then the
proper length of the $S^1$ of the boundary is  
\begin{equation}
  \beta = T^{-1} = \int_0^{\beta_{SGUP}} d\tau \sqrt{g_{\tau\tau}} =
  \beta_{SGUP} \sqrt{f(R)}. \label{beta}
\end{equation}
Since $\sqrt{\gamma}$ and $K$ are explicitly calculated as 
$\sqrt{\gamma} = R^2 \sin\theta \sqrt{f(R)}$, $K = - 2\sqrt{f(R)}/R -
M/R^2\sqrt{f(R)}$ at the boundary, and $I_1$ becomes
\begin{equation}
  I_1 = \beta_{SGUP} \left(\frac32 M - R\right). \label{I1:value}
\end{equation}
On the other hand, $I_{\rm 0}$ is evaluated for a flat four-metric
with boundary $S^1\times S^2$. In this case, the period of the
Euclidean time is $\beta$ and we have $\sqrt{\gamma} = R^2 \sin
\theta$ and $K = -2/R$, which yields $I_{\rm 0} = -\beta R$. This
subtraction term normalizes the thermal energy to zero for the
Schwarzschild geometry with $M=0$ and has no effect on the other
physical quantities. Combining these two terms, the Euclidean action
becomes
\begin{equation}
  \label{I}
  I = \beta R + \frac{\beta}{\sqrt{f(R)}} \left(\frac32 M - R\right).
\end{equation}

From Eqs.~(\ref{T*}) and (\ref{beta}), the local temperature measured
on the boundary in a thermal equilibrium is 
\begin{equation}
  \label{T}
  T = \frac{M}{4\pi\ell^2} \left( 1 - \sqrt{1 -
  \frac{4\ell^2}{L^2} - \frac{\ell^2}{M^2}} \right) \left(
  1-\frac{2M}{R} \right)^{-\frac12},
\end{equation}
which is nothing but the Tolman temperature with the redshift factor. 
The interesting point to distinguish from the conventional Tolman
temperature is that $M$ in Eq.~(\ref{T}) is bounded and all its value lies
between $\ell/\sqrt{1-4\ell^2/L^2} < M < R/2$.
Even at the minimal black hole, the temperature is finite
$T=1/4 \pi \ell\sqrt{1-4\ell^2/L^2}$ for $M=\ell/\sqrt{1-4\ell^2/L^2}$
while it is divergent for $M=0$ in the conventional Tolman temperature. 
 
Since the area of $S^2$ of the boundary is $A=4\pi R^2$, the total
thermodynamic internal energy within the boundary $R$ becomes
\begin{equation}
  \label{E}
  E = \left( \frac{\partial I}{\partial\beta} \right)_A = R -
  R\sqrt{1-\frac{2M}{R}} \left[\frac{1-\varepsilon-
    r_0/R}{1 - \varepsilon - r_c/R}\right],
\end{equation}
where $\varepsilon = 4(M/L)^2(1-4\ell^2/L^2)$, $r_0 = (3M/2)
(1-4\ell^2/L^2+\alpha )$, $r_c = M(2 - \varepsilon - 4\ell^2/L^2 +
\alpha)$, and $\alpha=\sqrt{1-4\ell^2/L^2 - \ell^2 / M^2}$. In the
finite $R$, $E \approx R - R(1-3M/2R)/\sqrt{1-2M/R}$ for which $M$
goes to $\ell/\sqrt{1-4\ell^2/L^2}$ and $E \approx 2M$ as $M$ goes to
$R/2$, respectively. What the nonvanishing black hole mass even in
this minimal black hole means is that there is a positive definite
smallest energy corresponding to a remnant. The thermodynamic energy
is singular at $r_c =R(1-\varepsilon)$ where this point corresponds to
the critical mass $M_c\equiv M|_{\partial  T/\partial M=0}$, on which
the temperature goes to the critical temperature $T_c\equiv
T|_{M=M_c}$. Note that below the critical temperature, no black
hole exists.

\begin{figure}[pt]
  \includegraphics[width=0.5\textwidth]{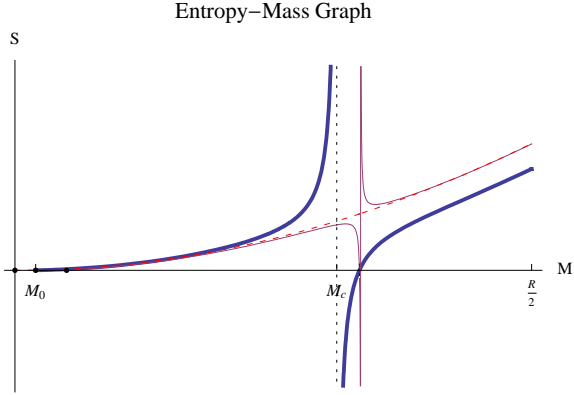}
  \caption{Three entropies are plotted for the case of $L=20$ and $\ell=1/5$ (SGUP;
    thick solid line), 
$L\to\infty$ and $\ell=1/2$ (GUP; thin solid line), 
and $L\to\infty$ and $\ell=0$ (HUP; dashed line). Each
entropy becomes zero at the corresponding minimal masses $M_0
=\ell/\sqrt{1-4\ell^2/L^2}$, $\ell$, and 0. 
The three points near the origin in the horizontal axis represent 
the minimal mass for HUP, SGUP, and GUP cases from the left, respectively.}
  \label{fig:entropy}
\end{figure}

From the free energy relation, $F=E-TS$, the black hole entropy is
explicitly written as
\begin{equation}
  \label{S}
  S = \beta E - I = 2\pi M^2 \alpha \left( 1-\frac{4\ell^2}{L^2} +
    \alpha \right) \frac{1-3M/R}{1-\varepsilon-r_c/R},
\end{equation}
where we used Eqs.~(\ref{I}), (\ref{T}), and (\ref{E}). 
Note that it can be reduced to the well-known area law $S= 4 \pi M^2$
for $\ell \rightarrow 0$ and $L\to\infty$ which is independent of the
size of the cavity so that it suggests that the black hole entropy is
independent of the asymptotic behavior of the gravitational field and
matter fields. However, once the minimal length is assumed, then the
entropy is related to not only the minimal length but also the
boundary through the cavity size. Apparently, the area law is no
longer hold, but it can be easily proved that the thermodynamic first
law, $dE = TdS$, is automatically satisfied for fixed $A$. It is
interesting to note that the entropy of the minimal black hole is zero
which means that the minimal black hole state whose mass is
$M=\ell/\sqrt{1-4\ell^2/L^2}$ can be a single state at the end of the
black hole evaporation. We plotted the entropies which correspond to
HUP, GUP, and SGUP in Fig.~\ref{fig:entropy}.

\section{GUP case of $L \rightarrow \infty$}
\label{sec:largeL}
The thermodynamic stability of the black hole can be studied in terms of
the heat capacity in the GUP limit. For this purpose, we consider a
large $L$ limit of $L\to\infty$, so that the temperature~(\ref{T}) is
reduced to
\begin{equation}
  T = \frac{M}{4\pi\ell^2} \left( 1 - \sqrt{1 - \frac{\ell^2}{M^2}}
  \right) \left( 1-\frac{2M}{R} \right)^{-\frac12}.
\end{equation}
\begin{figure}[pt]
  \includegraphics[width=0.5\textwidth]{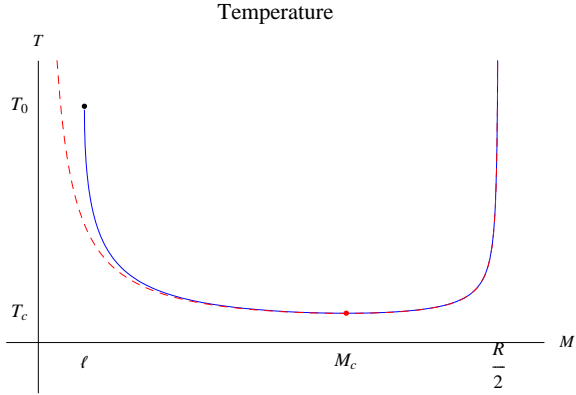}
  \caption{The solid line and the dotted line show the 
        profiles of the temperature based on the GUP and the HUP,
        respectively. The crucial difference between them
      comes near the end state of the black hole, and the minimal
    length prevents the black hole from the total evaporation
    [$\ell=1$, $R=20$ : $M_c \approx 6.69171$, $T_c \approx 0.010396$,
    $T_0 \approx 0.083882$].}
  \label{fig:T}
\end{figure}

In the conventional analysis with a cavity, the behavior of
temperature of a small black hole looks similar to that of the large
black hole since the small black hole has the small horizon
radius compared to the size of the cavity so that its temperature gets
large as in the HUP case, while the large black hole which is
comparable to the size of the cavity gives a high temperature due to
the redshift factor since the local observer is almost near the
horizon. However, in the present calculations based on the GUP, the
black hole does not completely evaporate, in other words, the
temperature of the small black hole is not so high because of the cutoff.

As seen in Fig.~\ref{fig:T}, there are no black hole states for
$T<T_c$, and 
both small and large black hole can exist within $T_c < T
< T_0$ where $T_0 = T|_{M=\ell} = \left( 4\pi \ell \sqrt{1 - 2\ell/R}
\right)^{-1}$, while only the large black hole solution is possible
for $T>T_0$. Note that there is a forbidden region between $0<M<\ell$,
which tells us that there is no small black hole whose mass is less
than the minimal length dimension. On the other hand, it is possible
to obtain the minimal black hole of $M=\ell$ in contrast to the
conventional thermodynamical analysis.

\begin{figure}[pt]
  \includegraphics[width=0.5\textwidth]{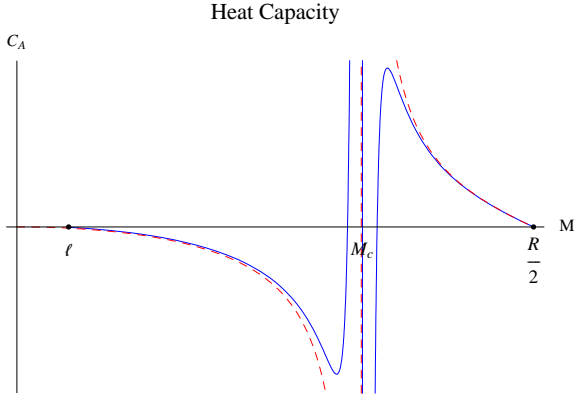}
  \caption{The dotted and the solid line show the behaviors of heat
    capacities based on the HUP and GUP, respectively.     
    This figure is plotted for $\ell=1$, $R=20$ : $M_c
    \approx 6.69171$, 
    $C_A|_{M\to\ell} \approx -5.93412$.}
  \label{fig:C}
\end{figure}
To study the thermodynamic stability of a black hole, one can consider
the heat capacity at a constant surface, which is defined by 
\begin{eqnarray}
  C_A &\equiv&
    \left( \frac{\partial E}{\partial T} \right)_A \nonumber \\
  &=& -2\pi M^2 \left( 1 - \frac{r_c}{M} \right)^2 \left(
    1-\frac{2M}{R} \right) \frac{(1-r_-/R)(1-r_+/R)}{(1-r_c/R)^3},
    \label{C} 
\end{eqnarray}
where the constants are $r_\pm = a \pm \sqrt{a^2-b}$, $a =
(M^3/\ell^2)[ 1 + 3 \ell^2 / (2M^2) - (1 - \ell^2 / M^2)^{3/2} ]$, 
and $b = 9M^2[1-\ell^2/(3M^2)]$, respectively. If we take the limit of
$\ell \to 0$, it is naturally reduced to the result of the
HUP~\cite{york},
\begin{equation}
  \label{C:HUP}
  C_A^{\rm HUP} = - 8\pi M^2 \left(1 - \frac{2M}{R} \right) \left(1 -
    \frac{3M}{R} \right)^{-1},
\end{equation}
which is positive for $R/3 < M < R/2$ while it is negative for
$0<M<R/3$. Moreover, it is singular at $M \rightarrow M_c=R/3$.
In this case, the small black hole of $M \rightarrow 0$ is unstable
to decay into either into pure thermal radiation or to a large
black hole. In our generalized case as shown in Fig.~\ref{fig:C}, 
there are two different aspects from the HUP case.
First, the critical behavior of the heat capacity  near the critical
mass is more complicated so that we can not obtain the definite
criteria for the stability for $M<M_c$ or $M>M_c$. For $M <M_c$($M
>M_c$), there may exist a stable(unstable) region near the critical mass
while most part of the region is unstable (stable). 
Secondly, the minimal black hole exists in our analysis so that the
black hole can not evaporate completely. For
$M \to \ell$ in Eq.~(\ref{C}), 
the heat capacity of the minimal black hole is $C_A \approx
-2\pi\ell^2$, where the horizon of the black hole is
the same with the minimal length, $r_{H}=2M=2 \ell =\Delta x_\textrm{min}$.

\section{Another case of finite $L$} 
\label{sec:comp}
In this section, we are now in a position to study how the SGUP
affects the thermodynamic quantities by assuming $L$ to be finite,
while, in the previous section, we have studied the GUP limit of $L
\rightarrow \infty$.
\begin{figure}[pt]
  \includegraphics[width=0.5\textwidth]{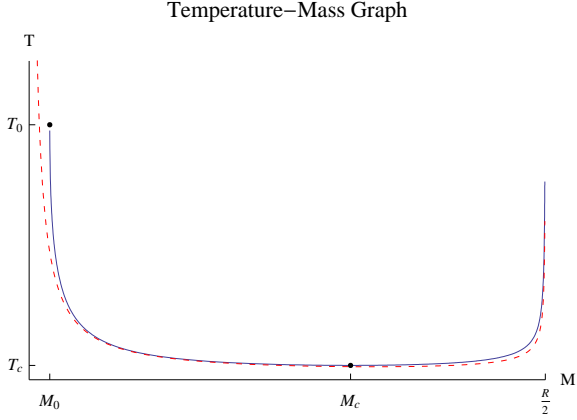}
  \caption{The solid line describes the behavior of the temperature for
    the finite $L$, which gives an additional correction mainly 
     to the large black hole. Eventually, the temperature are shifted
     up due to the two uncertainty constants $\ell$ and $L$.
 [$L=20$, $\ell=1/5$, $R=10$ : $M_c \approx
    3.11365$, $T_c \approx 0.0228477$, $M_0 \approx 0.20004$, $T_0 \approx 0.406175$].} 
  \label{fig:T:comparable}
\end{figure}
As seen in Fig.~\ref{fig:T:comparable}, the finite $L$ gives a
temperature correction mainly to the large mass black hole comparable
to the cavity size, $M\sim R/2$. Note that the minimal black hole mass
$M_0$ is larger than that of GUP so that it gives higher temperature.

We calculate the heat capacity at a constant surface to see whether a
black hole in a cavity is stable or not, which is now given by
\begin{equation}
  C_A = -2\pi M^2 \left( 1 - \frac{4\ell^2}{L^2} + \alpha \right)^2
  \left(1-\frac{2M}{R} \right) \frac{({\cal A}-2{\cal B}/R+{\cal
    C}/R^2)}{(1-\varepsilon-r_c/R)^3},   
\label{C:comp}
\end{equation}
where the constants are ${\cal A} = (1-4\ell^2/L^2) (1+4\alpha
M^2/L^2)$, ${\cal B} = (M^3/\ell^2) [(1-4\ell^2/L^2) (
1-\varepsilon+4(\ell/L)^2(1+5\alpha/2)) + (3/2-10\ell^2/L^2)
\ell^2/M^2 - \alpha^3(1-\varepsilon)]$, and 
${\cal C} = 9M^2(1-4\ell^2/L^2-\ell^2/3M^2)(1+4\alpha M^2/L^2)$,
respectively. If we take the limit of $L\to\infty$, then
it naturally recovers the heat capacity in Eq.~(\ref{C}).
On the other hand, as for the other extreme limit of $\ell \to 0$
keeping $L$ to be finite, the heat capacity becomes
\begin{equation}
  C_A = - 8\pi M^2 \left(1 - \frac{2M}{R} \right) \left( {\cal A} - 
\frac{2{\cal B}}{R} +\frac{{\cal C}^2}{R^2} \right) \left[1 -
\frac{4M^2}{L^2} - \frac{M}{R}\left(3-\frac{4M^2}{L^2}\right)
\right]^{-1},
\end{equation}
where ${\cal A}\to(1+4M^2/L^2)$, ${\cal B} \to M(3+10M^2/L^2-8M^4/L^4)$, 
and ${\cal C}\to9M^2(1+4M^2/L^2)$. Of course, the heat
capacity~(\ref{C:HUP}) of the HUP case can be derived for $L\to\infty$
and $\ell\to0$.
\begin{figure}[pt]
  \includegraphics[width=0.5\textwidth]{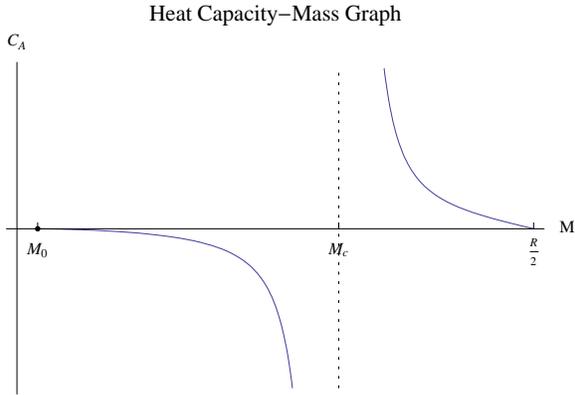}
  \caption{The heat capacity is singular at the critical mass and it
    is ill-defined when the mass is less than the minimal mass. It
    also shows that it is negative for $M < M_c$ while it is positive
    for $M > M_c$. This figure shows that it is critically different
    from that of the GUP case. The heat capacity goes to $C_A \approx
    -2\pi M_0^2(1-3M_0/R)/(1-2M_0/R)$ as $M \to M_0$ and zero as $M \to
    R/2$. This figure is plotted for $L=20$, $\ell=1/5$, $R=10$ : $M_c
    \approx 3.11365$, $M_0 \approx 0.20004$, $C_A|_{M\to M_0} \approx
    -0.174082$.}
  \label{fig:C:comparable}
\end{figure}
In Fig.~\ref{fig:C:comparable}, the complicated behavior of the heat
capacity near the critical mass disappears and resembles the behavior
of the heat capacity of HUP. Therefore, the whole profile of the heat
capacity for the finite $L$ case is very similar to that of the HUP
case. In fact, the complicated behavior of the heat capacity near the
critical mass appears again when both $L \gg 1/\ell$ and $L \gg R$
are satisfied. The exact critical condition exists but it is too
lengthy to write down.

\section{Discussion}
\label{sec:dis}
We have studied thermodynamic quantities and the stability of the
black hole in a cavity based on the extended GUP called symmetric GUP
whose limits are well defined such as $L \rightarrow \infty$ and $\ell
\to 0$. In particular, following the Euclidean action formalism in a
cavity, the Tolman temperature has been used in deriving the heat
capacity, and it gives the consistent thermodynamic first law along
with the appropriate energy and the entropy. Moreover, this entropy
can not be written in the form of the area, whereas it recovers the
area law when we take the HUP limit of $L \rightarrow \infty$ and
$\ell \to 0 $. The entropy is usually proportional to the area since
it is independent of the asymptotic behavior of the gravitational
field and matter fields; however, it depends on two uncertainty
parameters, $L$ and $\ell$ in our case.

In the HUP case, a small black hole in unstable equilibrium over the
critical temperature may decay either into pure thermal radiation or
to a large black hole. In the GUP case, the off-shell free energy
~\cite{gpy,hp,sh} can be defined by $F_\textrm{offsell}=E-TS$ where
Eqs.~(\ref{E}) and (\ref{S}) are used with the arbitrary
temperature. It shows that the small black hole in the unstable
equilibrium  may decay into the minimal black hole or a larger stable
black hole, however, whose final mass should be smaller than the
critical mass in Fig.~\ref{fig:T}. From the extrema condition of the
free energy, $(dF_\textrm{offshell}/dM)_{R}=0$, it can be shown that
the mass of the small unstable black hole can not exceed the critical
mass, because the off-shell free energy (potential) is singular at the
critical mass due to the definition of the GUP temperature of Eq.~(\ref{T}). 
If the initial mass of the black hole is larger than the critical
mass, then it may decay into a really large black hole whose mass is
nearly $R/2$, which is also a big difference from the HUP case. 

In fact, there remains further issues in connection with GUP(SGUP)
calculations. First, the Hawking temperature modified by the GUP in
Ref.~\cite{acs} should be confirmed by other independent methods. One
of them may be a scattering method in connection with the grey-body
factor. Second, the resulting modified temperature~(\ref{T*}) subject
to the GUP(SGUP) should be consistent with the metric giving the
well-known Hawking temperature through the periodicity of the
Euclideanized metric. It means that we have to consider the back
reaction of the geometry properly. We hope these issues will be
addressed elsewhere. 
 

\acknowledgments
This work was supported by the Sogang Research Grant, 20071063 (2007)
and the Science Research Center Program of the Korea Science and
Engineering Foundation through the Center for Quantum Spacetime
(CQUeST) of Sogang University with grant number R11-2005-021.


\end{document}